\documentclass[aps,prb,twocolumn,showpacs,preprintnumbers,superscriptaddress]{revtex4}
\usepackage{graphicx}
\usepackage{amssymb}
\usepackage{bm}
\usepackage{amsmath}
\usepackage{color}

\def\QE{{\sc Quantum ESPRESSO}}
\def\yambo{{\sc yambo}}

\begin{document}

\title{Probing optical excitations in chevron-like armchair graphene nanoribbons}

%

\author{Richard Denk}\thanks{These authors equally contributed}
\affiliation{Institute of Experimental Physics, Johannes Kepler University, 4040 Linz, Austria.}
\author{Alberto Lodi-Rizzini}\thanks{These authors equally contributed}
\affiliation{Department of Physics, Mathematics, and Informatics, University of Modena and Reggio Emilia, 41125 Modena, Italy.}
\affiliation{Centro S3, CNR--Istituto Nanoscienze, 41125 Modena, Italy.}
\author{Shudong Wang}\thanks{These authors equally contributed}
\affiliation{Centro S3, CNR--Istituto Nanoscienze, 41125 Modena, Italy.}
\author{Michael Hohage}
\affiliation{Institute of Experimental Physics, Johannes Kepler University, 4040 Linz, Austria.}
\author{Peter Zeppenfeld}
\affiliation{Institute of Experimental Physics, Johannes Kepler University, 4040 Linz, Austria.}
\author{Jinming Cai}
\affiliation{Empa, Swiss Federal Laboratories for Materials Science and Technology, 8600 Duebendorf, Switzerland, Department of Chemistry and Biochemistry, University of Bern, 3012 Bern, Switzerland.}
\author{Roman Fasel}
\affiliation{Empa, Swiss Federal Laboratories for Materials Science and Technology, 8600 Duebendorf, Switzerland, Department of Chemistry and Biochemistry, University of Bern, 3012 Bern, Switzerland.}
\author{Pascal Ruffieux}
\affiliation{Empa, Swiss Federal Laboratories for Materials Science and Technology, 8600 Duebendorf, Switzerland, Department of Chemistry and Biochemistry, University of Bern, 3012 Bern, Switzerland.}
\author{Reinhard Berger}
\affiliation{Center for Advancing Electronics Dresden and Department of Chemistry and Food Chemistry, Technische Universit\"at Dresden, Mommsenstra{\ss}e 4, D-01062 Dresden, Germany}
\author{Z. Chen}
\author{Akimitsu Narita}
\affiliation{Max Planck Institute for polymer research, 55128 Mainz, Germany.}
\author{Xiliang Feng}
\affiliation{Center for Advancing Electronics Dresden and Department of Chemistry and Food Chemistry, Technische Universit\"at Dresden, Mommsenstra{\ss}e 4, D-01062 Dresden, Germany}
\author{Klaus M\"ullen}
\affiliation{Max Planck Institute for polymer research, 55128 Mainz, Germany.}
\author{Roberto Biagi}
\affiliation{Department of Physics, Mathematics, and Informatics, University of Modena and Reggio Emilia, 41125 Modena, Italy.}
\affiliation{Centro S3, CNR--Istituto Nanoscienze, 41125 Modena, Italy.}
\author{Valentina De Renzi}
\affiliation{Department of Physics, Mathematics, and Informatics, University of Modena and Reggio Emilia, 41125 Modena, Italy.}
\affiliation{Centro S3, CNR--Istituto Nanoscienze, 41125 Modena, Italy.}
\author{Deborah Prezzi}
\email[corresponding author: ]{deborah.prezzi@nano.cnr.it}
\affiliation{Centro S3, CNR--Istituto Nanoscienze, 41125 Modena, Italy.}
\author{Alice Ruini}
\affiliation{Department of Physics, Mathematics, and Informatics, University of Modena and Reggio Emilia, 41125 Modena, Italy.}
\affiliation{Centro S3, CNR--Istituto Nanoscienze, 41125 Modena, Italy.}
\author{Andrea Ferretti}
\affiliation{Centro S3, CNR--Istituto Nanoscienze, 41125 Modena, Italy.}

\begin{abstract}
The bottom-up fabrication graphene nanoribbons (GNRs) has opened new opportunities to specifically control their electronic and
optical properties by precisely controlling their atomic structure. Here, we address excitations in GNRs with periodic
structural wiggles, so-called Chevron GNRs. Based on reflectance difference and high-resolution electron energy loss
spectroscopies together with ab-initio simulations, we demonstrate that their excited-state properties are dominated by
strongly bound excitons. The spectral fingerprints corresponding to different reaction stages in their bottom-up fabrication are also unequivocally identified, allowing us to follow the exciton build-up from the starting monomer precursor 
to the final ribbon structure. 
\end{abstract}

\date{\today}

\maketitle

\section{Introduction}

\begin{figure*}
\includegraphics[width=0.9\textwidth,clip]{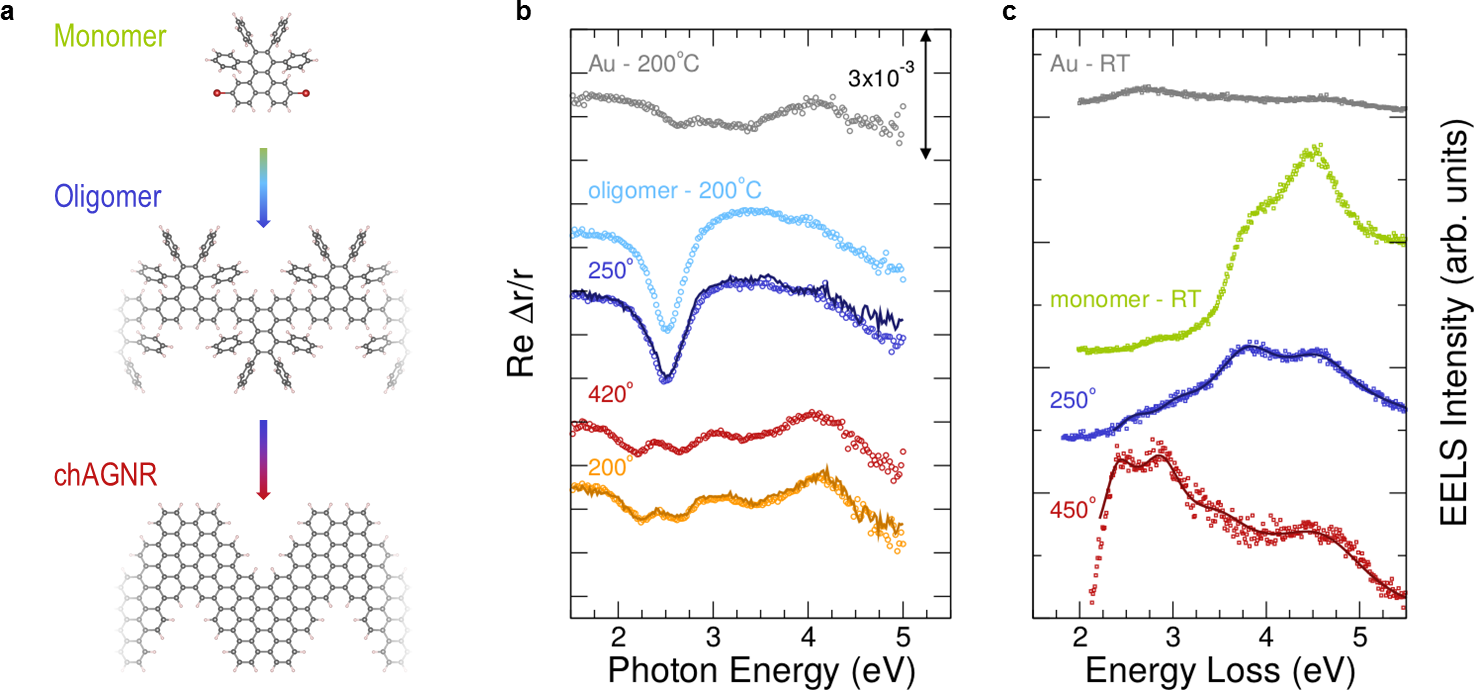}
\caption{(a) Sketch of the thermally-activated synthesis route
used to produce chevron-like AGNRs. (b) Re $\left(\Delta r/r\right)$ (RDS) and (c) EELS data taken at different stages of the reaction, during ch-AGNR growth: clean Au surface (grey symbols), monomer at RT (green squares), intermediate polymer (light blue and blue symbols), ch-AGNR (red and orange symbols).  The solid lines in (b) and (c) are fitted spectra based on the three-phase model and harmonic oscillator transitions. For better viewing, the spectra are vertically shifted.
\label{fig:growth-RDS-EELS}}
\end{figure*}

Successful fabrication of atomically precise graphene nanoribbons (GNRs) has boosted current research in view of possible applications in the fields of photonics and optoelectronics 
\cite{Soavi2016,Zhong2015,Zhong2017,Candini2017}. 
Despite the recent advancements, the investigation of the GNR optical properties
is still at the early stages \cite{Narita2014,Denk2014,Soavi2016,Zhao2017,Senkovskiy2017}.
Indeed, not only different edge types can lead to radically different properties, but also 
the ribbon shape as well as edge functionalization can play a significant role \cite{Tan2013,Narita2015,Ivanov2017}.
%
Here we focus on the case of chevron-like armchair GNRs (ch-AGNRs). These GNRs, 
synthesized~\cite{Cai_nature_2010} from 6,11-dibromo-
1,2,3,4-tetraphenyltriphenylene monomers,
are characterized by a wiggle-like structure and armchair-edged termination (see Fig.~\ref{fig:growth-RDS-EELS}a). 
Because of their geometry, these GNRs can have different properties with respect to straight armchair GNRs, e.g. in view of thermoelectric applications~\cite{sevi+13scirep,lian+12prb}. 
Moreover, their electronic properties can be easily tuned by edge-doping, as in the paradigmatic case 
of N substitution \cite{bron+13ac}: GNRs with different degrees of N-doping have been synthesized
starting from nitrogen-substituted monomers, and the resulting systems display the same configuration as pristine ch-AGNR, but shifted electronic bands.
This has also opened the way to their exploitation 
as bulding blocks for more complex nanostructures, 
such as tunable graphene-based type-II nanojunctions~\cite{Cai2014}, obtained by co-deposition of pristine and N-doped 
monomers on the same surface. In this perspective, 
ch-GNR can also form 
threefold junctions\cite{Cai_nature_2010},  which were fabricated 
by exploiting the C3 symmetry of tri-halogen-functionalized monomers.
 
Chevron-like AGNRs thus represent a very promising class of graphene-based nano-objects, 
whose electronic and optical properties deserve a thorough investigation.
While the  optical excitations of extremely low-dimensional systems are generally 
expected to be dominated by excitonic effects, this has been experimentally proven 
only for a limited number of cases. Among GNRs, pronounced excitonic effects
were demonstrated only for the case of ultranarrow (N=7) armchair 
ribbons on gold substrates~\cite{Denk2014} 
by exploiting their fully anisotropic 
optical properties and performing reflectance difference spectroscopy (RDS) measurements. 
In the case of wiggle-edged ch-AGNR, the 1D character (and associated quantum confinement) 
is less pronounced than for straight-edge ribbons (such as AGNRs) with similar width, 
thus challenging the prediction of large excitonic effects as well as 
their determination by means of anisotropy-exploiting spectroscopies. 
So far, either the interpretation of experimental findings 
for the case of ch-AGNRs  
was based on a single-particle picture, where optical 
and electronic gap are treated as the same concept, 
or excitonic fingerprints were not
recognized~\cite{Vo2014natcomm,Shekhirev2017,bron+13ac,Vo2015}.

 
In this paper, the excited-state properties of chevron-like GNRs
are investigated in a combined experimental and theoretical
study, with the aim of providing a comprehensive picture 
of their low-energy (UV-vis) excitations.
 Moreover, monitoring the  build-up of the optical excitations  
during   the  growth process,  allows us to   clearly identify 
the optical fingerprints of each sequential formation step. 
Experimentally, RDS and high-resolution
electron-energy loss spectroscopy (HREELS) are employed: the resulting 
 information are combined, and further supported 
by state-of-the-art ab initio calculations based on many-body perturbation 
theory (MBPT) methods, such as $GW$ and Bethe Salpeter equation (BSE).
%


We find that the complementary
information inherent to RDS and HREELS results can be 
well-interpreted within the framework of the dielectric function tensor theory, 
showing  excellent quantitative agreement. Furthermore, the experimental results 
can be rationalized by means of MBPT calculations, revealing unambiguously that 
the response of ch-AGNR ribbons is exciton-dominated at UV-vis frequencies, 
with exciton binding energies larger than 1 eV, as previously found for narrower 
ribbons~\cite{Denk2014}. 

\section{Methods}
 %
\subsection{RDS measurements} 
%
RDS is an optical method sensitive to the in-plane optical anisotropy of a sample \cite{Weightman2005}. 
It measures the real and imaginary parts of the normalized difference of the complex reflection coefficients $r$ at normal incidence for light polarized along two orthogonal axes $x$ and $y$ in the surface plane~\cite{Weightman2005}: 
\begin{equation}
  \frac{\Delta r}{r}=2\frac{r_x-r_y}{r_x+r_y}.
\end{equation}
%
%
The RD spectrometer (ISA Jobin Yvon) used in this study allows for recording of the RD signal in the energy range between 1.5 eV and 5.0 eV. A strain-free quartz viewport (Bombco Inc.) mounted on the ultrahigh vacuum system provides the optical access to the sample at normal incidence.

The ch-AGNRs were grown on the regularly stepped Au(788) surface to achieve a high degree of unidirectional alignment of the nanoribbons required for RDS \cite{Linden2012}, as previously done for 7-AGNRs \cite{Ruffieux2012}.
The ch-AGNR fabrication steps were followed by recording RD spectra of the pristine Au(788) sample at 150$^{\circ}$C and 200$^{\circ}$C, after monomer deposition at 200$^{\circ}$C, during annealing at 250$^{\circ}$C and 420$^{\circ}$C, and once more at 200$^{\circ}$C after the annealing.  The RD spectrometer was aligned such that the $x$ and $y$ directions were parallel and perpendicular to the GNR axis, and thus corresponded to the $[01\bar{1}]$ and $[\bar{2}11]$ axes of the Au(788) sample, respectively.
  
\subsection{HREELS measurements} 
%
In constrast to RDS, system anisotropy is not a requirement for HREELS-based investigations of the dielectric properties so that a Au(111) surface was employed as a template for the ribbon growth.
On this surface, GNR are known to grow with essentially no preferential orientation.
The  ch-AGNR formation steps were characterized by  HREEL vibrational spectra, in combination with XPS, which provided information on overlayer stoichiometry and coverage (see the Supplementary Information for details). The coverage of the monomer precursor phase corresponds to $\sim$ 2ML (1ML is defined as a single molecular layer with surface number density of $6.4\times10^{13}$ molecules/cm$^2$).
 
Instead, for the on-surface synthesis of the polymer and the ch-AGNR phases, 
0.5 ML of the monomer were deposited at RT on a freshly-cleaned  surface and subsequently annealed for 10 minutes  at 250$^{\circ}$C and  450$^{\circ}$C, respectively.
For each fabrication step, electronic spectra (1-6 eV energy loss region) were recorded  in specular geometry, with a primary beam energy $E_p$=9 eV and an energy resolution of 15 meV. 


\subsection{First-principles simulations}\label{sec:method_theo}
%
The electronic and optical properties of ch-AGNRs and their polymer precursors were simulated within an ab-initio many-body perturbation theory scheme~\cite{onid+02rmp}. 
We first computed the equilibrium atomic positions and electronic ground state by using a total-energy-and-forces approach based on density functional theory (DFT), pseudopotentials, and plane-waves, as implemented in the \QE{} package \cite{gian+09jpcm}. We employed the LDA exchange-correlation functional and norm-conserving pseudopotentials, with 45 Ry energy cutoff on the wavefunctions. A vacuum region of about 15 \AA\, in the non-periodic directions was introduced to prevent interaction between periodic replicas. Atomic positions were fully relaxed until forces were smaller than 0.0005 Ry/Bohr (0.013 eV/\AA).

Next, we simulated the quasiparticle band structure within the $GW$ approximation to the electron self-energy (in the $G_0W_0$ approximation 
within the plasmon-pole model~\cite{godb-need89prl}).  
The calculation of the macroscopic absorption spectrum was performed by 
solving the Bethe-Salpeter equation (BSE) for the $GW$-corrected quasi-electrons and quasi-holes (within the Tamm-Dancoff approximation, 
whose validity was explicitly verified 
for quite similar systems~\cite{prez+08prb}).
A truncation scheme~\cite{rozz+06prb} for the Coulomb potential was adopted 
to avoid spurious interactions between replicas. 
Both $GW$ and BSE calculations were performed with the \yambo{} code~\cite{mari+09cpc}.

Concerning GW/BSE calculations, the Brillouin zone is sampled by $22\times1\times1$ ($14\times1\times1$) $\mathbf{k}$-points for ch-AGNRs (precursor polymer).
The sum-over-states for ch-AGNRs (precursor polymer) in the calculation of polarization function and Green function have been truncated at 400 (400) and 450 (350) bands, respectively. The optical absorption spectra are calculated including 15 (6) valence bands and 15 (6) conduction bands. The kinetic energy cutoff to represent the response functions of GW and BSE for the ch-AGNRs [precursor polymer] corresponds to 2000 reciprocal lattice vectors (RL) ($\sim$4.5 Ry) [2600 RL ($\sim$1.5 Ry)] and 3000 RL ($\sim$7 Ry) [4000 RL ($\sim$2 Ry)], respectively.

\section{Results and discussion}\label{sec:results}

\subsection{Monitoring the growth process}

Figure~\ref{fig:growth-RDS-EELS} displays the evolution of the real part of the RD (b) and the EELS (c) spectra during the ch-AGNR on-surface synthesis (see Supplementary Information for the imaginary part of the RD spectra). A cartoon of the main steps of the process is shown in panel (a). As detailed in Sec.~\ref{sec:3phase}, RDS and EELS provide different but complementary information. Indeed, RDS records the differential reflection coefficient $\Delta r/r$ for polarization along two perpendicular directions of the sample, thereby emphasizing the dielectric anisotropy and suppressing isotropic contributions. 
From this, it is possible to extract the differential dielectric function $\Delta\epsilon = \epsilon_{x}-\epsilon_{y}$.
In contrast, due to the random molecular orientation on Au(111), 
the dielectric function extracted from HREELS data is an average, i.e.  $\bar{\epsilon} = (\epsilon_{x}+\epsilon_{y})/2$ (where $x$ and $y$ again refer to the polarization directions parallel and perpendicular to the polymer/GNR axis, respectively).
%
The finite RD signal recorded from the pristine Au(788)  substrate 
[gray dots in Fig.~\ref{fig:growth-RDS-EELS}(b)]
arises only from the optical anisotropy of its topmost layers, with no contribution from  the optically isotropic bulk. Instead, the EELS signal [gray squares in Fig.~\ref{fig:growth-RDS-EELS}(c)] stems from  the isotropic Au(111) substrate.
In the same way,  the optically isotropic  monomer adlayer is almost featureless in RDS (not shown here), while EELS is able to unequivocally capture its optical response, characterized by two intense features in the region above 3.5 eV [green squares in Fig.~\ref{fig:growth-RDS-EELS}(c)].


Upon heating the sample to 250$^{\circ}$C, 
the EELS spectrum clearly changes, pointing to the formation of a new phase~\cite{Cai_nature_2010} [Fig.~\ref{fig:growth-RDS-EELS}(c), blue squares].
The high-energy features are still present, although with different relative intensity ({\it i.e.}, some spectral weight is transferred to the lower energy state). In addition, a broad shoulder appears in the region below 3.5 eV, which is not present in the monomer phase. 
Insight is provided by the RD spectra recorded at 200$^{\circ}$C and 250$^{\circ}$C, which reveal a prominent feature below 3 eV, indicating that the new phase is anisotropic. This can be attributed to the formation of extended and aligned polymers, obtained by the concatenation of individual monomers.
The higher energy features observed in EELS  are instead not seen in RDS, and could be related to electronic states that are  
more isotropic, thus resulting in a much smaller signal in RDS.
The temperature increase from 200$^{\circ}$C to 250$^{\circ}$C leaves the RD spectrum essentially unchanged, suggesting that either the polymerization process is already completed at 200$^{\circ}$C, or the polymers are sufficiently long for the saturation of the optical properties to occur. 

By further increasing the temperature to 420$^{\circ}$C, the RD signal changes upon ch-AGNRs formation, with the appearance of additional features and an overall decrease of the intensity, the latter indicating a much less pronounced anisotropy of the ch-AGNRs as compared to the polymers. Indeed, the RD amplitudes for the ch-AGNRs are of the same size as the signal from the clean Au(788) substrate. 
In such a situation, a careful data analysis becomes mandatory for a sound interpretation of the results (see Sec. \ref{sec:3phase} and Supporting Information). 
The ch-AGNRs formation is also evident from the EEL spectra, where most of the spectral weight is transferred to the low-energy region below 3.0 eV. We here highlight that the spectra of both the polymer and the ch-AGNR phase are rather different from those previously reported in the literature\cite{bron+13ac}, where the polymer and the GNR spectra  (taken at $E_p=15$ eV) were  both characterized by a rather broad peak at about 2.7-2.8 eV. In fact, this feature derives from the Au surface plasmon (SP) loss, which dominates the spectra and hinders a reliable determination of adsorbate-related features when recorded with high primary electron beam energy $E_p$~\cite{Vattuone2013}. We show in the Supplementary Information that an energy $E_p$ as low as 9~eV is needed to clearly distinguish the adsorbate features from those of the substrate.

So far we have demonstrated that both RDS and EELS are effective in monitoring the on-surface synthesis of GNRs, and provide complementary information. However, in order to make our analysis quantitative, we have to extract and compare from both datasets the same fundamental/physical quantity, namely, the dielectric function. In particular, in the case of monolayer-thin adsorbates, one needs to carefully single out the effect of the substrate to obtain information on the dielectric properties of the adsorbate. In the following section we briefly introduce the three-phase model~\cite{McIntyre1971,Cole1998,Weightman2005,ibac-mill82book} (3PM) that will be subsequently applied in the analysis of both RDS and HREELS data to extract the adsorbate dielectric function.

%
%

\subsection{Data analysis: three-phase model}
\label{sec:3phase}
%
%
According to the 3PM, we consider a thin adlayer with (possibly anisotropic) complex dielectric function $\epsilon(\omega)$ that is adsorbed on a semi-infinite bulk with dielectric function $\epsilon_b(\omega)$. The system is in touch with an ambient medium, in our case vacuum ($\epsilon_a=1$).
Moreover, we assume that the adsorbate is well decoupled from the surface, meaning that we can just subtract the Au signal to extract $\epsilon$ (or $\Delta\epsilon$) of the adsorbate.
In the following, the components of $\epsilon(\omega)$ are written as a sum of Lorentzian
oscillators,
\begin{equation}
   \epsilon_i(\omega)  = 1 + \sum_n \frac{ A_{n,i}}{E_{n,i}^{2}-\omega^2 -i \omega\Gamma_{n,i}},
   \label{eq:lorentz}
\end{equation}
($i$ standing for the in-plane polarization directions $x$ and $y$). The quantities $\{A_{n,i} ,E_{n,i}, \Gamma_{n,i}\}$ are then used as fitting parameters to reproduce the experimental data.


For ultrathin adlayers the complex reflection anisotropy of a three-layer system can be written in terms of the dielectric functions entering the 3PM~\cite{McIntyre1971,Cole1998,Weightman2005,Denk2014}
\begin{equation}
    \frac{\Delta r}{r} = 2\frac{ r_x-r_y}{r_x+r_y} = \frac{4\pi i d}{\lambda} \, 
        \frac{\Delta \epsilon}{\epsilon_b -1},
        \label{eq:RDS3PM}
\end{equation}
where $\Delta \epsilon = \epsilon_{x} -\epsilon_{y} $ is the dielectric anisotropy, $d$ is the thickness of the adsorbed layer (set for both RDS and EELS to 3 \AA{}), and $\lambda$ the wavelength of the incident light ($\lambda \gg d$).
In contrast to EELS (see below) the dielectric functions used here are  evaluated at $\mathbf{q}\to0$, i.e., zero momentum transfer. 
Using Eq.~(\ref{eq:RDS3PM}) it is possible to directly extract the dielectric anisotropy $\Delta \epsilon$ of the adsorbed layer from an RDS measurement ($\Delta r/r$), if the
dielectric function $\epsilon_b$ of the underlying substrate is known.
In view of the weak optical anisotropy of chevron-like GNRs (as compared to 7-AGNRs~\cite{Denk2014}), we have carefully checked the 3PM fitting procedure and the possible effect of different Au dielectric functions (see Supporting Information). As a result, we find no relevant changes in the fitted data. For the final analysis, we have chosen to subtract the substrate signal recorded at 150$^{\circ}$C and the gold dielectric function was taken from au\_2 of
WVASE32 database by J.A. Wollam Co., Inc, as in our previous analysis of 7-AGNRs in Ref.~\cite{Denk2014}.

Concerning EELS, in the dipole-scattering approximation, the measured spectra are proportional to the system loss function 
\begin{equation}
    L(\omega,\mathbf{q}) = -\text{Im} \left[
       \frac{1}{\epsilon_t(\omega,\mathbf{q})+1}
    \right].
\label{eq:eels1}
\end{equation}
The loss function $L$ depends on the dielectric function of the overall system $\epsilon_t(\omega,\mathbf{q})$, which can be written, according to the 3PM~\cite{ibac-mill82book}, as a weighted average of the adsorbate and  bulk dielectric functions $\bar{\epsilon}$ and $\epsilon_b$, respectively:
\begin{eqnarray}
    \epsilon_t(\omega,\mathbf{q}) &=& \bar{\epsilon}(\omega,\mathbf{q})\, 
         \frac{1+D(\omega,\mathbf{q})\,e^{-2qd}} {1- D(\omega,\mathbf{q})\,e^{-2qd}}, \\
    D(\omega,\mathbf{q}) &=& 
    \frac{\epsilon_b-\bar{\epsilon}}{\epsilon_b+\bar{\epsilon}},
\label{eq:eels2}
\end{eqnarray}
where  $d$ is the thickness of the adsorbate layer, and $\mathbf{q}$ the momentum transfer parallel to the surface plane. 
The system loss function [with $\bar{\epsilon}$ parameterized by a sum of Lorentzian oscillators as in Eq.~(\ref{eq:lorentz})] can thus be obtained if the substrate dielectric function is known. For direct comparison with the RDS results we use again the tabulated dielectric function au\_2 of
WVASE32 database by J.A. Wollam Co., Inc. 
The validity of our model was first checked by comparing the calculated and experimental EEL spectra for the case of the clean gold substrate. Full details are provided in the Supporting Information. 


\subsection{Dielectric properties of ch-AGNRs and polymer precursors}

\begin{figure*}
\includegraphics[width=0.9\textwidth]{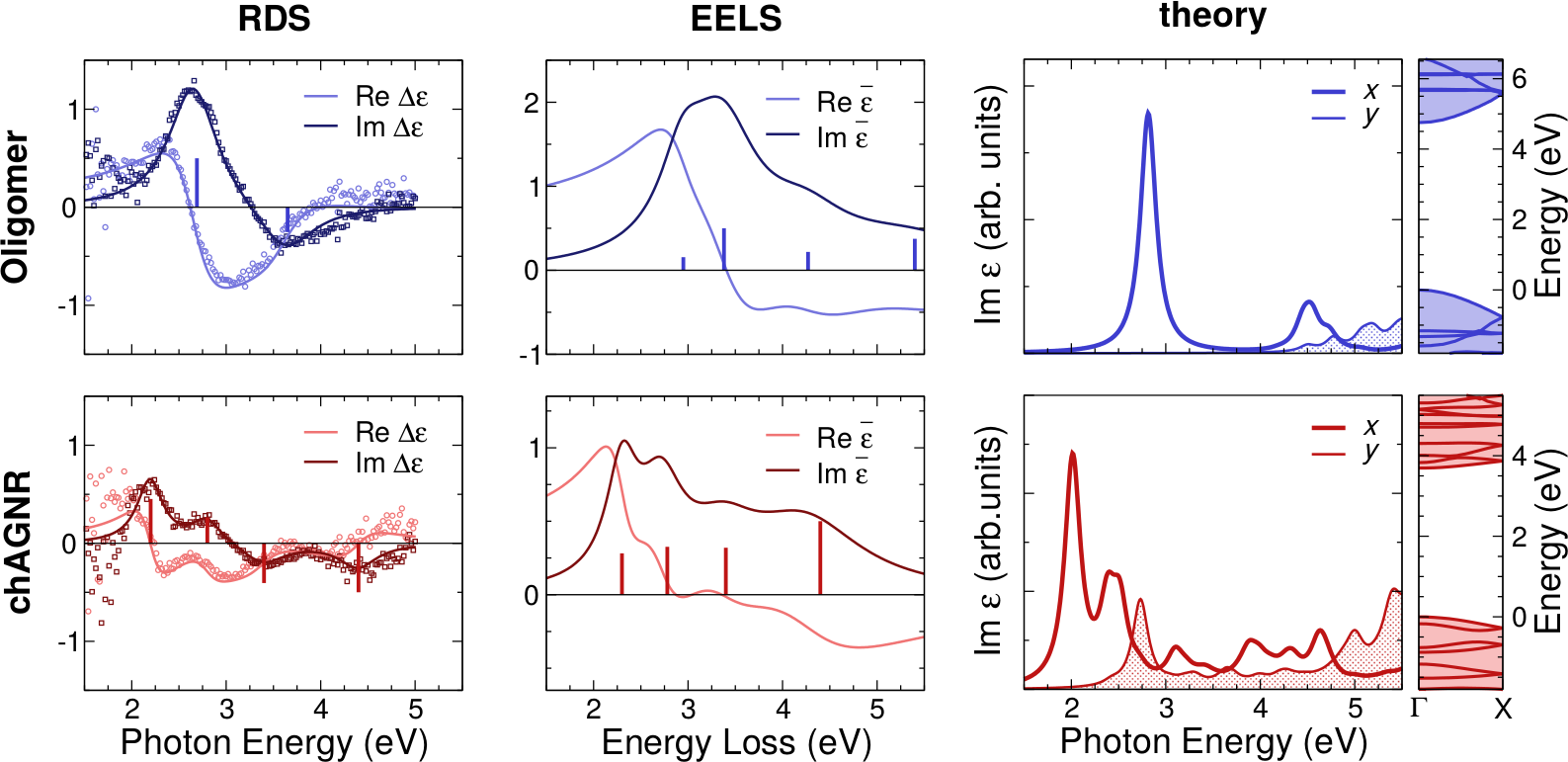}
\caption{Dielectric functions of precursor polymer (blue) and ch-AGNR (red) extracted from the RDS data according to Eq.~\ref{eq:RDS3PM} (symbols) and fitted to a Lorentzian oscillator model (lines) (left panels), computed from the EELS data (Eqs.~\ref{eq:eels1}-\ref{eq:eels2}) (central panels) data, and computed by BSE (right panels). Positions and intensities of the fitted Lorentzian oscillator transitions are shown as vertical bars.
\label{fig:epsilon}}
\end{figure*}

\begin{figure}
\includegraphics[width=0.45\textwidth]{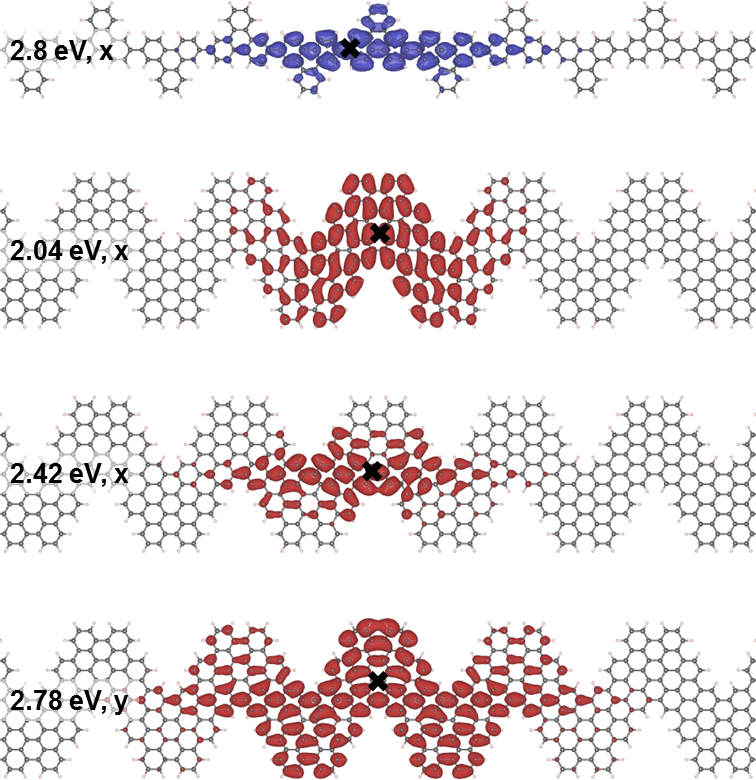}
\caption{Real-space,  electron-hole probability distribution for selected bright excitons of the precursor polymer (blue) and the ch-AGNR (red). 
The fixed hole position is 
marked with a black cross.
The exciton energies and the light polarization are indicated in the figure: ($x$) light polarized along the polymer/ribbon axis (longitudinal polarization) ($y$) light polarized in-plane and perpendicular to the polymer/ribbon axis (transverse polarization)
  \label{fig:excwfc}}
\end{figure}

\begin{table}
  \caption{Parameters of the Lorentzian oscillators used to parameterize $\Delta\epsilon$ and $\bar{\epsilon}$ for modeling the RDS and EELS  spectra of the polymer and ch-AGNR adlayer. Symbols are defined in Eq.~(\ref{eq:lorentz}). The last column reports the peak positions in the calculated optical absorption spectrum.
  	 \label{tab:optics_data}}
\begin{ruledtabular}
  \begin{tabular}{ccccccccc}
  \multicolumn{3}{c}{\textbf{RDS ($\Delta\epsilon$)}} & & \multicolumn{3}{c}{\textbf{HREELS ($\bar{\epsilon}$)}} &
  & {\textbf{Theory ($\epsilon_x$, $\epsilon_y$)}}\\
  \hline
   $A_n$   &  $E_n$   & $\Gamma_n$  & &  $A_n$  &  $E_n$  & $\Gamma_n$ & &  $E_n$ \\ 
    & (eV)  & (eV) & &  & (eV)  & (eV) & & (eV)\\[5pt]
  \hline
  \multicolumn{9}{l}{\textbf{polymer}}\\
  \hline
   2.74      &   2.69   &  0.68 & &  1.5 & 2.95  & 0.58 & & 2.7 ($x$) \\
   -1.39      &   3.65   &  0.74 & &  4.8 & 3.38  & 0.92 & & 4.5 ($y$) \\
   -    &   -   &  -  & &  2.1 &  4.27 & 1.0   & & \\
   -      &   -   &  -  & &  3.6  &  5.4 & 2.0  & & \\
  \hline
  \multicolumn{9}{l}{\textbf{ch-AGNR}}\\
  \hline
    0.48 &  2.20   &  0.34 & &  1.41 &  2.30 & 0.40 & & 2.04 ($x$) \\
    0.28 &  2.80   &  0.40 & &  1.63 &  2.78 & 0.54 & & 2.40 ($x$) \\
   -0.43 &  3.40   &  0.59 & &  1.6  &  3.40 & 0.90  & & 2.80 ($y$) \\
   -0.53 &  4.40   &  0.50 & &  2.5  &  4.40 & 1.0  & &
  \end{tabular}
  \end{ruledtabular}
\end{table}

Figure~\ref{fig:epsilon} compares the dielectric anisotropy of the adlayer ($\Delta\epsilon$) obtained from the RDS data to the average dielectric function of the adlayer ($\bar{\epsilon}$) extracted from the EEL spectra, following the data analysis described in Sec. \ref{sec:3phase}. The $\Delta\epsilon_2$ spectrum of the polymer precursor extracted from the RDS data is dominated by a positive peak at $\sim$2.7 eV (absorption of light polarized along the polymer axis $x$), and a weaker negative peak around 3.6 eV (absorption of light perpendicular to the polymer axis). $\Delta\epsilon$ can be readily extracted from the measured spectra using Eq.~(\ref{eq:RDS3PM}) (the peak at 2.7~eV even without subtraction of the Au signal).
The fit to a Lorentzian oscillator model (Eq.~(\ref{eq:lorentz}), solid lines) reveals that two transitions are sufficient to reproduce $\Delta\epsilon$ (see Table~\ref{tab:optics_data}). This compares well with the average dielectric function $\bar{\epsilon}$ for the polymer obtained from EELS, which is reproduced by considering one sharp excitation, located  at 2.9 eV, and three broader features at 3.4, 4.3 and 5.4 eV, respectively. 
While the low-energy features agree within 0.2 eV, the higher energy ones are much less pronounced in $\Delta\epsilon$, indicating that they are related to less anisotropic excitations.  


%
Overall, the above analysis nicely demonstrates the consistency of the RDS and EELS data.
The differences in the oscillator intensities are due to the fact that different quantities ($\Delta\epsilon$ and $\bar{\epsilon}$) are measured, while
the small  discrepancies in the excitation energies can be accounted for by taking into consideration the $\mathbf{q}$-dependence of the EELS loss-function. $\bar{\epsilon}$ as determined by EELS compares also very well to an absorbance peak at 3.2 eV reported for dispersed polymers~\cite{Vo2014natcomm}.

The results of the 3PM analysis are also in agreement with the low-energy region of the optical absorption spectrum computed from first principles, which is dominated (for the longitudinal polarization $x$) by a single, prominent peak at 2.7 eV (see upper right panel in Fig.~\ref{fig:epsilon}) mainly originating from the HOMO to LUMO transitions around the $\Gamma$-point of the Brillouin zone. This spectral feature is excitonic in nature. The real-space analysis of the electron-hole probability distribution shows a strong one-dimensional localization across the polymer backbone (top plot in Fig.~\ref{fig:excwfc}, the hole position is fixed and indicated by a dark spot), while it is delocalized over a few monomer units along the polymer axis (Wannier-like exciton). Evaluating the extension of the e-h distribution at 90$\%$ of its height, we find a value of $\sim$ 35~\AA{}(corresponding to about 4 monomer units). This provides an estimate for the saturation lenght~\cite{Varsano2008prl} of the low-energy optical feature: since the RD spectrum recorded from the polymer phase is already saturated at 200$^{\circ}$C (see Fig.~\ref{fig:growth-RDS-EELS}(b)), we can conclude that a significant fraction of the molecules has already reacted into short polymers.
%
Concerning the higher energy excitations, we unfortunately cannot comment on their nature due to the approximations chosen to address this structure. In particular, the phenyl rings decorating the polymer backbone were pruned to make the calculations more treatable (see Sec.~\ref{sec:method_theo}). As a consequence, excitations which involve 
the outer phenyl rings are absent in the present calculations. 
It is very likely, that the transition at about 3.5 eV seen both 
in RDS and EELS but not reproduced by theory is exactly such 
a  phenyl ring related  transition  (see also the feature above 3.5 eV
in the spectrum shown in Fig.~\ref{fig:growth-RDS-EELS}(c) green symbols) 

When the chGNRs are formed from the polymer precursors, we notice a clear change in the RD spectrum, from which we can extract four main  transitions that contribute to $\Delta\epsilon$: two of them, located at 2.2 and 2.8 eV, respectively, are for light polarized along the GNR axis ($x$), while the other two, at 3.4 and 4.4 eV, involve light polarized along $y$, i.e. perpendicular to the GNR axis. 
The EELS-derived dielectric function $\bar{\epsilon}$ for the ch-AGNR is characterized by two main low-energy peaks, located at $\sim$2.3 and 2.8 eV, respectively; two additional features at higher energies ($\sim$3.4 and 4.4 eV) can also be recognized, showing again a remarkable agreement between RDS and EELS results (the transition energies agree within 0.1 eV; the EELS features are slightly broader than the RDS peaks). 

The calculated optical spectrum of ch-AGNRs (lower right panel in Fig.~\ref{fig:epsilon}) shows a similar pattern as the $\Delta\epsilon$ function from RDS: two low-energy peaks for light polarized along the ribbon axis, located at 2.04 (mainly derived from HOMO to LUMO transitions) and 2.42 eV (HOMO-1 to LUMO+3), respectively; a first peak at 2.80 eV (HOMO to LUMO+1) for light polarized perpendicular to the ribbon axis, and a large number of excitations of both polarizations at higher energy. While the low energy features are in good agreement with RDS data, the higher energy ones show larger discrepancies. However, this is expected from our approximation where the substrate is totally neglected. It is  interesting to compare the electron-hole probability distribution of the ch-AGNRs with that of the polymer (Fig.~\ref{fig:excwfc}). The first exciton at 2.04~eV has similar extension along the axis than on to the polymer (at 2.8~eV), but the localization in the perpendicular direction is much less pronounced. On the contrary, the second exciton of the ch-AGNR at 2.42~eV is slightly more delocalized along the axis, but thinner in the perpendicular direction. Its energy and shape are remarkably similar to the lowest excitation of the polymer system.

\section{Conclusions}
In conclusion, we have investigated the optical excitations 
in chevron-like GNRs by means of a joint experimental-theoretical study, 
where RDS and HREELS are combined with ab initio calculations 
based on many-body perturbation theory. 
The excellent agreement of results obtained through two different,
complementary spectroscopies, and their successful comparison with theoretical data  
(see Tab.~\ref{tab:optics_data}), allow us to shine light on 
the optical response of ch-AGNRs and their precursors, also sorting out previous contradictory interpretations.
%
%
Firstly, we show that taking into account the substrate contribution to the measured spectra is mandatory for a correct interpretation of the experimental data. 
Moreover, our analysis confirms that excitons are primary excitations in ch-AGNRs, leading to electron-hole binding energies
of about 1.5 eV for self-standing ribbons.
Finally, the distinct optical signatures recorded
for the intermediate polymer precursors and the resulting ribbons 
suggest that RDS and EELS  
are effective tools in monitoring the growth and catching the ch-AGNRs in the act of their formation.

\end{document}